\documentclass[journal]{IEEEtran}
\ifCLASSINFOpdf
\else
\fi
\usepackage{amssymb}
\usepackage{graphicx}
\usepackage{amsmath}
\usepackage{multirow}
\newcommand{\ignore}[1]{}
\newif\ifarxivversion
 \arxivversiontrue  

\ifarxivversion
\pdfoutput=1
\fi
\begin{document}
%
\title{Gradient-Domain Fusion for Color Correction in Large EM Image Stacks}
%
%
%

\author{Michael~Kazhdan, Kunal~Lillaney, William~Roncal,\\ Davi~Bock, Joshua~Vogelstein, and~Randal~Burns
\IEEEcompsocitemizethanks{%
\IEEEcompsocthanksitem M. Kazhdan, K. Lillaney, W. Roncal, and R. Burns are with the Department of Computer Science, Johns Hopkins University.
\IEEEcompsocthanksitem D. Bock is with Janelia Farms, Howard Hughes Medical Institute.%
\IEEEcompsocthanksitem J. Vogelstein is with the Institute for Computational Medicine, Johns Hopkins University.}%
\thanks{}}

%
%

\markboth{Gradient-Domain Fusion for Color Correction in Large EM Image Stacks
}%
{Gradient-Domain Fusion EM Image-Processing}
%



\maketitle

\begin{abstract}
We propose a new gradient-domain technique for processing registered EM image stacks to remove inter-image discontinuities while preserving intra-image detail. To this end, we process the image stack by first performing anisotropic smoothing along the slice axis and then solving a Poisson equation within each slice to re-introduce the detail. The final image stack is continuous across the slice axis and maintains sharp details within each slice. Adapting existing out-of-core techniques for solving the linear system, we describe a parallel algorithm with time complexity that is linear in the size of the data and space complexity that is sub-linear, allowing us to process datasets as large as five teravoxels with a 600 MB memory footprint.
\end{abstract}

\begin{IEEEkeywords}
Gradient Domain, Image Processing, Image Fusion
\end{IEEEkeywords}

\ifCLASSOPTIONpeerreview
\begin{center} \bfseries EDICS Category: TEC-RST \end{center}
\fi
%
\IEEEpeerreviewmaketitle

\section{Introduction}
\label{s:intro}
Recent innovation and automation of electron microscopy sectioning has made it possible to obtain high-resolution image stacks capturing the relationships between cellular structures~\cite{Hayworth:MM:2006}. This, in turn, has motivated research in areas such as  connectomics~\cite{Anderson:BP:2009,Biswal:PNAS:2010,Bock:N:2011,Anderson:MV:2011} which aims to gain insight into neural function through the study of the connectivity network.

\begin{figure}[h]
\includegraphics[width=\columnwidth]{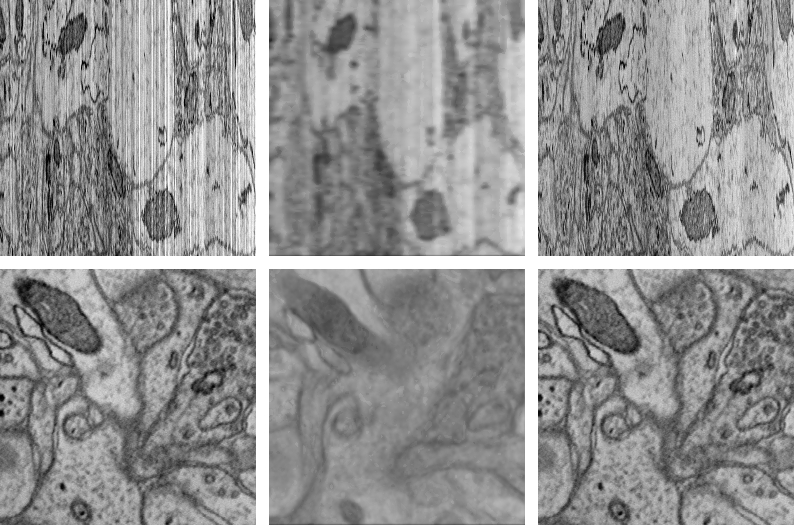}
\caption{
\label{f:data}
Cross-sections of an EM stack showing $zy$-slices through the data (top) and $xy$-slices through the data (bottom). The image on the left is taken from the original data, the image in the center is the result of the initial anisotropic smoothing step, and the image on the right is the subsequent solution of the screened-Poisson equation.
}
\end{figure}

While the technological advances in acquisition and registration have made it possible to acquire unprecedentedly large micron-resolution volumes, the acquisition process itself introduces undesirable artifacts in the data, complicating tasks of (semi-)automatic anatomy tracking. Specifically, since the individual slices in the stack are imaged independently, discontinuities often arise between successive slices due to variations in lighting, camera parameters, and the physical manner in which a slice is positioned on the slide. An example of these artifacts can be seen in Figure~\ref{f:data} (top-left), which shows an image of the same column taken from successive images in a stack (1850 images at a resolution of $21,\!504\times26,\!624$) imaging a mouse cortex~\cite{Kat11}. The visualization highlights the local discontinuities (thin vertical stripes across the image) that can arise due to the acquisition process.

In this work we propose a new gradient-domain technique for processing these anatomical volumes to remove the undesired artifacts. The processing consists of two phases. In the first phase, we perform anisotropic smoothing across the slice axis to smooth out the discontinuities between these slices. As Figure~\ref{f:data} (center) shows, this has the desirable effect of removing the discontinuities (top) but it also smooths out the anatomical features within the slice (bottom). To address this, we perform a second step of gradient-domain processing on each slice independently, solving a screened-Poisson equation to generate a new voxel grid with low-frequency content taken from the anisotropically smoothed grid and high-frequency content taken from the original data. As  Figure~\ref{f:data} (right) shows, this combines the best parts of both datasets -- like the anisotropically smoothed grid, this solution does not exhibit discontinuities between slices (top), while simultaneously preserving the sharp detail present in the original data (bottom).

Our implementation of the gradient-domain processing is enabled by adapting an existing, out-of-core Poisson solver~\cite{DMG} to support the frequency-based merging of two images. As a result, our implementation is parallelizable, has linear time complexity, and sub-linear space complexity, supporting the efficient processing of truly large datasets.

\section{Related Work}
\label{s:related}
Over the last decade, gradient-domain approaches have gained prevalence in image processing~\cite{Agrawal:ICCV:2007}. Examples include removal of light and shadow effects~\cite{Horn:CGIP:1974,Finlayson:ECCV:2002}, reduction of dynamic range~\cite{Fattal:SIGGRAPH:2002,Weyrich:SIGGRAPH:2007}, creation of intrinsic images~\cite{Weiss:ICCV:2001}, image stitching~\cite{Perez:SIGGRAPH:2003,Agarwala:SIGGRAPH:2004,Levin:ECCV:2004}, removal of reflections~\cite{Agrawal:SIGGRAPH:2005}, and gradient-based sharpening~\cite{Bhat:ECCV:2008}. 

The versatility of gradient-domain processing has led to the design of numerous methods for solving the underlying Poisson problem in the context of large 2D images, including adaptive~\cite{Agarwala:SIGGRAPH:2007}, and hierarchical~\cite{Kazhdan:SIGGRAPH:2008,Kazhdan:TOG:2010} solvers.

In this work we show that (1)~the problem of removing inter-slice discontinuities in EM images can be reduced to the problem of performing frequency-based fusion of individual image slices, and (2)~solving the the frequency-based fusion problem amounts to solving a 2D Poisson equation. This allows us to leverage efficient, out-of-core, and distributed Poisson solvers to process huge EM images.

This extends our earlier (publicly available but unpublished) research~\cite{Kazhdan:ArXiv:2013} by replacing the computationally expensive 3D Poisson solver with a simple blurring operator.

\section{EM Image Processing}
\label{s:gradient-domain}
The goal of our EM image processing is to remove the artifacts that arise due to the independent imaging of the slices in the 3D volume. In practice, the independent imaging results in a 3D volume that does not exhibit obvious artifacts when $xy$ slices are considered individually, but exhibits distracting ``popping'' artifacts when viewed across the slice axis. Considered in the frequency domain, the artifacts are low-frequency in the $xy$-direction (hence unnoticed when slices are considered individually) but high-frequency in the $z$-direction (hence the unwanted ``popping'').

If, for simplicity, we consider a 3D image as comprised of four components:
\begin{enumerate}
\item {\bf LL}: low $xy$ frequency and low $z$ frequency,
\item {\bf HL}: high $xy$ frequency and low $z$ frequency, 
\item {\bf LH}: low $xy$ frequency and high $z$ frequency, and
\item {\bf HH}: high $xy$ frequency and high $z$ frequency.
\end{enumerate}
the goal is to generate the signal where the {\bf LH} component is removed. We address this in two steps, first, we smooth across the slice direction to remove the {\bf LH} and {\bf HH} components and then we fuse back in the high-frequency content within each slice to get back the {\bf HH} component.

\subsection{Anisotropic Smoothing}
The smoothing of the image data is performed by convolving the 3D image with an anisotropic Gaussian, aligned with the coordinate axes, which has low variance in the $xy$-plane and larger variance in the $z$-direction:
$$I^1 = I^0 * G_{\sigma_{xy},\sigma_z}$$
with $G_{\sigma_{xy},\sigma_z}$ the anisotropic Gaussian:
$$G_{\sigma_{xy},\sigma_z}(x,y,z)=\frac{1}{\sigma_{xy}^2\sigma_z(\sqrt{2\pi})^3}e^{-(x^2+y^2)/2\sigma^2_{xy}}\cdot e^{-z^2/2\sigma^2_z}.$$

\subsection{Screened-Poisson Blending}
On its own, anisotropic smoothing dampens both the {\bf LH} and {\bf HH} components, removing desirable high-frequency content within a slice. This is visualized in Figure~\ref{f:data} (middle) -- the smoothing effectively removes the inter-slice discontinuities (top), but it also blends out the details within each image (bottom). This motivates a second processing stage in which we generate the slices of the new 3D image, $I^2$, by fusing the low-frequency data from the slices of $I^1$ and high-frequency data from the slices of $I^0$.

Conceptually, this can be implemented by iterating through the slices of $I^0$ and $I^1$ and performing a frequency-space blend to generate the slices of $I^2$, giving less weight to the data from $I^0$ at lower frequencies and more weight at higher frequencies. As we show below, this can be formulated as a set of 2D gradient-domain problems where we seek a 3D image whose $j$-th slice minimizes:
$$I^2_j = \mathop{\arg\,\min}_I\int_{\Omega_j}
\alpha\left(I-I^1_j \right)^2 + \left\|\nabla I-\nabla I^0_j\right\|^2 dp.$$
Here $\Omega_j$ is the slice domain and $\alpha$ is the screening weight balancing the importance of interpolating pixel values of $I^1_j$ with the goal of matching the gradients of $I^0_j$.
Using the Euler-Lagrange formulation, the minimizer is obtained by solving the linear (Screened-Poisson) system:
\begin{equation}
\label{eq:screened_poisson}
(\alpha-\Delta) I^2_j = \alpha I^1_j - \Delta I^0_j
\end{equation}
for each slice $j$.

As visualized in Figure~\ref{f:data} (right), the second step of processing re-introduces the {\bf HH} component, providing a 3D image that has the sharp intra-slice detail of the input, $I^0$, without the inter-slice discontinuities. (More precisely, this step fuses back in the {\bf HL} and {\bf HH} components of $I^0$, but since the {\bf HL} component is already in $I^1$, only the {\bf HH} component changes.)

\subsection*{Frequency-Space Interpretation}
As in the work of Bhat~{\em et al}.~\cite{Bhat:ECCV:2008}, we get a frequency-space interpretation of Screened-Poisson blending by considering the system in the Fourier domain. Specifically, expressing the 2D slices $I_j^0$ and $I_j^1$ in terms of their frequency decomposition:
\begin{align*}
I_j^0(x,y)&=\sum_{k,l}\hat{I}_j^0(k,l)e^{2\pi i ( k x + l y)}\\
I_j^1(x,y)&=\sum_{k,l}\hat{I}_j^1(k,l)e^{2\pi i ( k x + l y)}
\end{align*}
and using the fact that the complex exponential $e^{2\pi i ( k x + l y)}$ is an eigenvector of the Laplace operator with eigenvalue $-4\pi^2(k^2+l^2)$ we get:
$$\hat{I}^2_j(k,l) = \frac{\alpha\hat{I}^1_j(k,l) + 4\pi^2(k^2+l^2)\hat{I}^0_j(k,l)}{\alpha+4\pi^2(k^2+l^2)}.$$

Thus, the $(k,l)$-th Fourier coefficient of $I^2_j$ is a weighted average of the $(k,l)$-th coefficients of $I^0_j$ and $I^1_j$, with the coefficient of $I^0_j$ receiving higher weight when the interpolation weight ($\alpha$) is small or the frequency ($k^2+l^2$) is large.

\subsection*{Implementation}
An advantage of our formulation is that it provides a simple and scalable solution for EM image processing.

The first step of our processing, anisotropic smoothing, is implemented by using a compactly supported approximation of the Gaussian (setting the weight to zero beyond three standard deviations). Because of the locality of the smoothing, this step can be implemented in a streaming fashion by maintaining a small window on the image in working memory, computing the weighted average for the in-core subset of the image (in parallel), and then advancing the window. This step has time complexity that is linear in the number of voxels in the 3D image and space complexity that only depends on the variances of the Gaussian, $\sigma_{xy}$ and $\sigma_z$.

The second step of our processing, solving the Poisson equation, is implemented by adapting the parallel/distributed and out-of-core solver of Kazhdan~{\em et al.}~\cite{Kazhdan:TOG:2010}. We modified the implementation in~\cite{DMG} to allow a user to input both a low- and a high-frequency image, setting up the constraints for the linear system as in Equation~\ref{eq:screened_poisson}, and then using the existing multigrid solver to obtain the solution. Each solve has time complexity that is linear in the number of pixels in a slice and space complexity that is linear in the size of the row.

Thus, for a 3D image with $O(N^3)$ voxels, our processing has time-complexity $O(N^3)$, space complexity $O(N)$, and can be implemented in parallel. Furthermore, our implementation depends on only three parameters -- the variances of the Gaussian used for anisotropic smoothing, $\sigma_{xy}$ and $\sigma_z$, and the interpolation weight used for frequency-based fusing, $\alpha$.

\paragraph*{Distributed Processing}
While our evaluations are performed on a single machine, our approach is also trivial to distribute. In the first phase we can leverage the fact that we use a compactly supported approximation of the Gaussian for smoothing, so computing the values in slice $I_j^1$ only requires knowing the values in slices $I_k^0$, with $|j-k|\leq 3\cdot\sigma_z$. In the second phase the 2D screened-Poisson equation is solved for each slice independently. Thus, the processing of $N$ slices can be distributed across $M$ machines by copying $N/M+6\cdot\sigma_z$ slices to each machine, and then computing the smoothed ($I^1$) and fused ($I^2$) values for the $N/M$ interior slices.


\subsection*{Alternate Solutions}
In addressing the color correction problem, we considered two other implementations.

In our initial research~\cite{Kazhdan:ArXiv:2013} we implemented the anisotropic smoothing by solving a gradient-domain problem in which the target gradient field was defined by zeroing out the $z$-components of the gradients of the input. We have opted for the simpler Gaussian convolution presented in this work because it is significantly faster in practice and has space complexity that depends on the size of the blurring kernel. In contrast, the gradient-domain implementation of anisotropic diffusion requires the implementation of a streaming 3D Poisson solver and has space complexity proportional to the size of an image slice.

We had also considered implementing the Poisson solver using the Fast Fourier Transform~\cite{Cooley:MC:1965,FFTW3:IEEE:2005}. However, we did not pursue this approach because the theoretical complexity of the FFT is slower than that of multigrid (log-linear vs. linear) and a scalable implementation requires an out-of-core transpose (often causing an I/O bottleneck for computation~\cite{Toledo:EMA:1999}).

\subsection*{Extensions and Applications}
\begin{figure}[h]
\begin{center}
\includegraphics[width=\columnwidth]{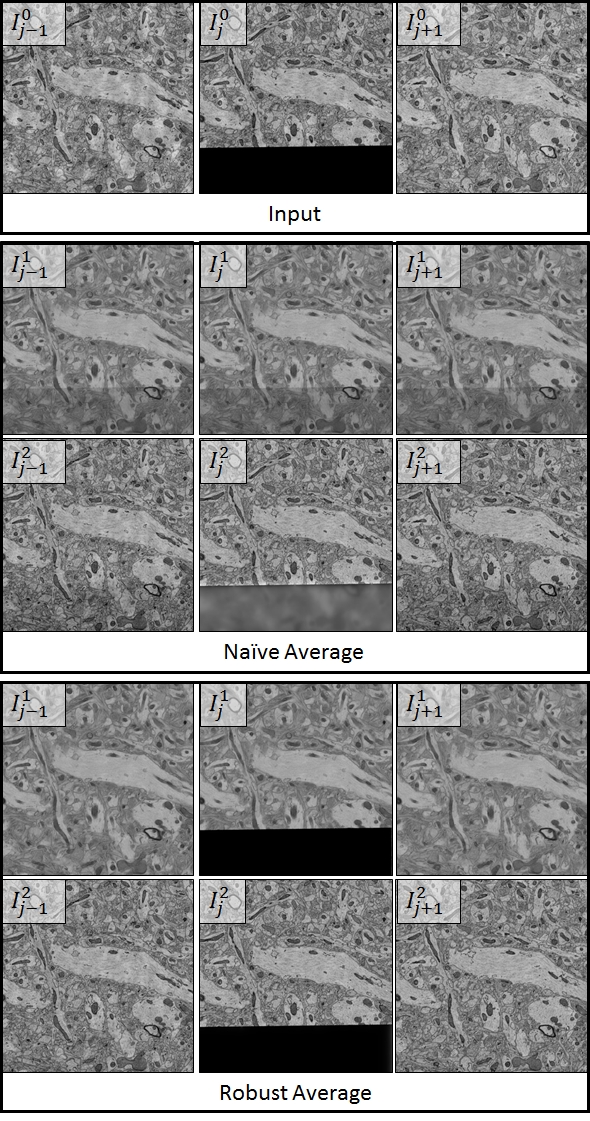}
\end{center}
\caption{
\label{f:outliers}
Comparison of naive (middle) and robust (bottom) averaging for successive slices $\{j-1,j,j+1\}$ from a 3D volume, showing the input data, $I^0$, the output from the smoothing phase, $I^1$, and the result of frequency-based fusion, $I^2$. By ignoring outlying color values when computing the weighted average across slices, we obtain a target low-frequency signal that mitigates the effects of bad data.
}
\end{figure}

An advantage of our formulation is that our decomposition of the computation into two simple steps makes it easy to adapt the processing. As an example, Figure~\ref{f:outliers} demonstrates how robust averaging can be incorporated to mitigate the effects of bad data by showing three successive slices in a volume, where the middle slice of the input is missing data (top row). Using naive smoothing distributes the corrupt data into adjacent low-frequency slices (second row) which then introduces artifacts in the output, visible as a darker band in the lower part of the neighboring slices (third row). Instead, by only averaging color values within two standard deviations, we obtain robust low-frequency slices (fourth row) that avoid introducing artifacts into adjacent output slices and do not introduce a low-frequency solution into the region with missing data (bottom row).

\begin{figure}[h]
\begin{center}
\includegraphics[width=0.7\columnwidth]{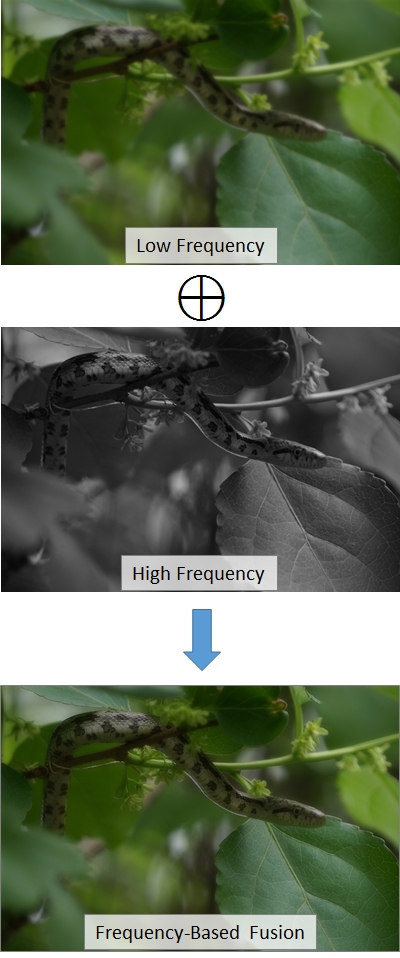}
\end{center}
\caption{
\label{f:snakes}
Frequency-based fusion for merging high-frequency monochrome images (top left) with low-frequency color images (bottom left). By using the color image to define the low frequency components and the monochrome for the high-frequency, we obtain a new image that successfully incorporates the color and detail from both (right).
}
\end{figure}

Additionally, though this work focuses on the processing of EM image stacks, we believe that our formulation of frequency-based fusion in the gradient domain contributes an general-purpose tool for image processing. As an example, Figure~\ref{f:snakes} shows an application of fusing a low-frequency color image with a high-frequency monochromatic image. Using the color image to define the value constraints and the monochromatic image to define the gradient constraints, the solution to the screened-Poisson equation provides a result that has both color and high frequency detail.\footnote{To define the gradient constraints, we obtained an RGB image by copying the gray-scale value into each of the three color channels.}

\section{Results}
\label{s:results}
\ignore
{
\noindent{\bf Notes}: What do we want to describe here?
\begin{itemize}
\item Running-time and memory for ac3, ac4, Kasthuri11, and Cardona.
\item Comparison to EMISAC on the ac3 and ac4 datasets: Running time and memory usage, side-by-side comparison, as well as quantitative evaluation.
\end{itemize}

\noindent{\bf Data}:
\begin{itemize}
\item ac3: $1024\times1024\times256\approx2.7\times10^8$.
\item ac4: $1024\times1024\times100\approx1.0\times10^8$.
\item Kasthuri11: $21,\!504\times26,\!624\times1850\approx1.1\times10^{12}$.
\item Cardona: $32,\!768\times32,\!768\times4840\approx5.2\times10^{12}$.
\end{itemize}
}

To evaluate our approach, we consider both the quality of the image and run-time performance. In these evaluations, we fix the radius of the in-slice blurring kernel to $\sigma_{xy}=1$, the cross-slice blurring kernel to $\sigma_z=3$ (both measured in pixels), and the screening weight to $\alpha=0.001$.

\subsection{Comparison to EMISAC}
We compare our method to the approach of Azadi~{\em et al.}~\cite{Azadi:ECCV:2014} (EMISAC). Similar to our original gradient-domain formulation~\cite{Kazhdan:ArXiv:2013}, EMISAC performs the editing by solving for the 3D image whose partial derivatives in the $x$- and $y$-directions match those of the input, and whose partials along the $z$-direction are close to zero. Unlike our earlier approach, EMISAC formulates the filtering as a solution to a constrained quadratic optimization problem, requiring the use of the computationally more expensive L-BFGS-B solver~\cite{Zhu:1997:TMS}.

We evaluate the two methods on a set of cut-outs from the ``Kasthuri11'' dataset~\cite{Kat11} of a mouse cortex.

\begin{figure}[h]
\begin{center}
\includegraphics[width=\columnwidth]{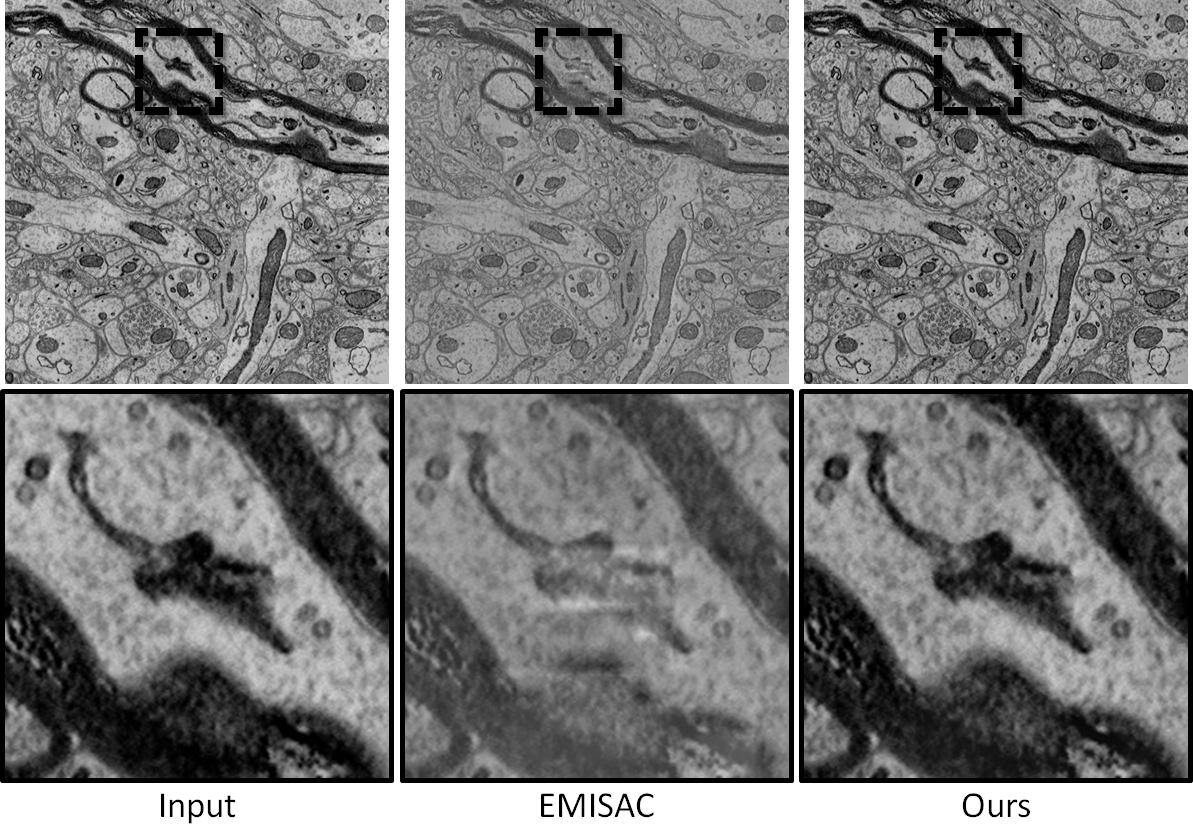}
\end{center}
\caption{
\label{f:EMISAC}
Visual comparison of our method with EMISAC: Showing a slice from the input (left), the corresponding slice in the results of EMISAC (center), and the corresponding slice in our results (right).
}
\end{figure}

Figure~\ref{f:EMISAC} compares the visual quality of the processed images which shows an $xy$-slice from the $1024\times1024\times100$ dataset. The image on the left shows a slice from the input, the one in the middle shows the corresponding slice after EMISAC filtering, and the one on the right shows the results of our approach. As can be seen from the figure, the EMISAC result has less contrast than the input and exhibits blocking artifacts.

Table~\ref{t:kasthuri11} compares the performance of the two approaches on the cut-outs, obtained on a machine with two Xeon E5630 @ 2.53 GHz processors and 64 GB of RAM, parallelized across eight threads. As the table shows, EMISAC's dependence on the L-BFGS-B solver comes at a noticeable increase in running time and memory. In the implementation of EMISAC, this is ameliorated  by decomposing the 3D volume into cells and solving the problem on each cell independently. However, replacing a global system with a set of local systems can introduce inaccuracies and may be the cause of the blocking artifacts noted above.

In contrast, the simplicity of our formulation results in a significantly more efficient implementation, with running times growing linearly with data size and in-core memory usage growing sub-linearly.

\ignore
{
\begin{table}[h]
\center{
\begin{tabular}{l|c|c|r@{ / }r|r@{ + }r|}
& \multicolumn{2}{c|}{EMISAC} & \multicolumn{4}{c|}{Ours} \\
\cline{2-7}
& \multicolumn{1}{c|}{Memory} & \multicolumn{1}{c|}{Time} &\multicolumn{2}{c|}{Memory} &\multicolumn{2}{c|}{Time} \\
\hline
ac3 & 39 (GB) & 4534 (s) & 64 & 50 (MB) & 37 & 168 (s) \\
\hline
ac4 & 16 (GB) & 1425 (s) & 64 & 50 (MB) & 15 & 41  (s) \\
\hline
\end{tabular}
}
\caption{
\label{t:ac3_ac4}
Performance comparison of our method with EMISAC: Showing peak memory usage and running time for processing the $1024\times1024\times256$ ``ac3'' dataset and $1024\times1024\times100$ ``ac4'' dataset.
}
\end{table}
}

\begin{table*}[h]
\center{
\begin{tabular}{l|r|c|r@{ / }r|r@{ + }l|}
& \multicolumn{2}{c|}{EMISAC} & \multicolumn{4}{c|}{Ours} \\
\cline{2-7}
& \multicolumn{1}{c|}{\multirow{2}{*}{Peak (MB)}} & \multicolumn{1}{c|}{\multirow{2}{*}{Time (h:mm:ss)}} &\multicolumn{2}{c|}{Peak (MB)} & \multicolumn{2}{c|}{Time (h:mm:ss)} \\
& & & Smooth & Blend & Smooth & Blend \\
\hline
$1024\times1024\times100$ &  6,577 & 0:13:59 &  59 & 39 & 0:00:24 & 0:00:55 \\
\hline
$1024\times1024\times200$ & 12,988 & 0:37:47 &  63 & 38 & 0:00:37 & 0:01:51 \\
\hline
$2048\times1024\times200$ & 26,089 & 1:16:59 &  68 & 34 & 0:01:23 & 0:01:53 \\
\hline
$2048\times2048\times200$ & 52,965 & 3:25:08 &  80 & 55 & 0:02:38 & 0:02:45 \\
\hline
$2048\times2048\times400$ &      * &       * & 105 & 57 & 0:05:18 & 0:05:26 \\
\hline
$4096\times2048\times400$ &      * &       * &  150 &  66 & 0:11:09 & 0:08:00 \\
\hline
$4096\times4096\times400$ &      * &       * &  202 & 123 & 0:21:22 & 0:13:07 \\
\hline
\end{tabular}
}
\caption{
\label{t:kasthuri11}
Performance comparison of our method with EMISAC: Showing peak memory usage and running time for processing different sized cutouts from the Kasthuri11 dataset. Peak memory usage and running time are provided separately for the smoothing and blending phases of our processing. (*At resolutions finer than $2048\times2048\times200$ the EMISAC implementation ran out of memory and could not complete.) 
}
\end{table*}

\subsection{Scalability}
To evaluate the scalability of our method, we also evaluated the performance on two large datasets. The datasets included the complete $1.2$ teravoxel ``Kasthuri11'' dataset as well as the $5.2$ teravoxel ``Cardona'' dataset from the Open Connectome Project~\cite{OCP}.

The performance of our approach on these two datasets is shown in Table~\ref{t:kasthuri_cardona}, obtained on a machine with two Xeon X5690 @ 3.47 GHz processors and 48 GB of RAM, parallelized across twelve threads. As the table shows, despite the large size of the datasets, the in-core memory usage remains negligible, due to the out-of-core implementation of the smoothing and blending phases. The table also confirms that the running time scales linearly with the resolution. (By comparison, when run on the ``Kasthuri11'' dataset, the first phase of our original gradient-domain implementation~\cite{Kazhdan:ArXiv:2013} took 41 hours and used 42 gigabytes of memory on the same machine.)

Combined with a parallelizable implementation, our approach provides a solution for processing registered EM image stacks that scales to the resolution of today's large datasets.

\begin{table*}[h]
\center{
\begin{tabular}{c|l|r@{ / }r|r@{ + }l|}
&  & \multicolumn{2}{c|}{Peak (MB)} & \multicolumn{2}{c|}{Time (h:mm:ss)} \\
& \multicolumn{1}{c|}{Resolution} & Smooth & Blend & Smooth & Blend \\
\hline
Kasthuri11 & $21,\!504\times26,\!624\times1850$ & 369 & 480 &  23:34:17 & 13:52:30 \\
Cardona    & $32,\!768\times32,\!768\times4840$ & 593 & 440 & 109:13:50 & 69:22:24 \\
\hline
\end{tabular}
}
\caption{
\label{t:kasthuri_cardona}
Performance of our method on two teravoxel datasets: Peak memory usage and running time are provided separately for the smoothing and blending phases of our processing.
}
\end{table*}

\section{Conclusion}
\label{s:conclusion}
This work analyzes the artifacts commonly arising in EM stacks due to the independent imaging of the slices. We propose a simple approach that first smooths the image along the slice axis and then performs frequency-based fusion to obtain an image that maintains the sharp detail within the slices while removing the discontinuity artifacts across them. We describe an extension of the well-established gradient-domain processing paradigm that implements the fusion by solving a Poisson equation, thereby providing a scalable parallel solution that has linear time and sublinear space. Finally, we demonstrate the effectiveness of the approach in processing images as large as five teravoxels using less than a gigabyte of working memory.


\ifCLASSOPTIONcaptionsoff
  \newpage
\fi



%
\bibliographystyle{IEEEtran}
\bibliography{paper}

%

\ignore
{
\begin{IEEEbiographynophoto}{John Doe}
Biography text here.
\end{IEEEbiographynophoto}


\begin{IEEEbiographynophoto}{Jane Doe}
Biography text here.
\end{IEEEbiographynophoto}
}




\end{document}